\documentclass[twocolumn,showpacs,amsmath,amssymb,pra,superscriptaddress,floatfix]{revtex4-1}

\usepackage{longtable}
\usepackage{amssymb, graphicx,subfigure}
\usepackage{enumerate}
\usepackage{nicefrac}
\usepackage{amsmath,bm}
\usepackage{hyperref}
\usepackage{epsfig}

\newcommand{\degree}{\ensuremath{^\circ}}%
\newcommand{\eg}{e.\,g.}%
\newcommand{\Magabs}{\ensuremath{\text{B}}}%
\newcommand{\Estatabs}{\ensuremath{\text{E}_{\textup{s}}}}%
\newcommand{\ie}{i.\,e.}%

\newcommand{\ind}[1]{_{\text{#1}}}

\newcommand{\ket}[1]{|#1\rangle}

\newcommand{\expected}[1]{\left\langle #1\right\rangle}

\begin{document}


\title{ Fine Structure of Open Shell Diatomic Molecules in Combined Electric and Magnetic Fields}

\author{Martin G\"arttner}\thanks{ {\em{Max Planck Institut f\"ur Kernphysik, Saupfercheckweg 1,
69117 Heidelberg, Germany}}}
\affiliation{Instituto Carlos I de F\'{\i}sica Te\'orica y Computacional,
and Departamento de F\'{\i}sica At\'omica, Molecular y Nuclear,
  Universidad de Granada, 18071 Granada, Spain}
\author{Juan J.\ Omiste}
\affiliation{Instituto Carlos I de F\'{\i}sica Te\'orica y Computacional,
and Departamento de F\'{\i}sica At\'omica, Molecular y Nuclear,
  Universidad de Granada, 18071 Granada, Spain}
\author{Peter Schmelcher}
\affiliation{Zentrum f\"ur Optische Quantentechnologien, Universit\"at Hamburg,
Luruper Chaussee 149, Hamburg, 22761, Germany}
\author{Rosario Gonz\'alez-F\'erez}
\email{rogonzal@ugr.es}
\affiliation{Instituto Carlos I de F\'{\i}sica Te\'orica y Computacional,
and Departamento de F\'{\i}sica At\'omica, Molecular y Nuclear,
  Universidad de Granada, 18071 Granada, Spain}
\date{\today}

\begin{abstract}

We present a theoretical study of the impact of an electric field combined with a magnetic field 
on the rotational dynamics of open shell diatomic molecules. Within the rigid rotor approximation,
we solve the time-independent Schr\"odinger equation including the fine-structure interactions
and the $\Lambda$-doubling effects. 
We consider three sets of molecule specific parameters and  several field regimes and  investigate
the interplay between the different interactions identifying  the dominant one.  The
possibility of inducing couplings between the spin and rotational degrees of freedom is demonstrated. 
\bigskip

\end{abstract}

\maketitle

\section{Introduction}

The control and manipulation of all molecular degrees of freedom, i.e., the center of mass,
electronic, rotational and vibrational motion, is an ambitious goal in modern molecular physics.
Ensembles of cold and ultracold molecules present a paradigm in this context. 
Within the last decade, many experimental techniques, based on the use of external 
fields, have been developed to create samples of cold and ultracold molecules
\cite{rempe2012,arxiv2012:ye,ANIE200805503,ni08,arxiv2012:ye_4069}. 
Currently, different species of cold  molecules are becoming available
\cite{PhysRevLett.109.115302,PhysRevLett.109.115303,PhysRevA.85.012511,PhysRevA.83.023418} 
with a special focus on heteronuclear alkali 
dimers~\cite{ni08,PhysRevLett.109.085301,C1CP21769K,PhysRevA.81.043637,deiglmayr:133004}.  
These experimental efforts should be accompanied by theoretical studies to understand how 
external fields modify the internal structure of these systems. 

In the recent years, a series of studies of the rotational spectrum of several 
diatomic molecules in a $^1\Sigma$ electronic state exposed to combined electric and magnetic
fields were performed~\cite{PhysRevA.78.033434,PhysRevA.79.013401}.
These systems are characterized by two distinct energy scales
associated to the rotational degrees of freedom and the hyperfine structure.
The next key structural ingredient is given by the electronic spin and orbital angular momentum, \ie,
molecules in a $^2\Sigma$ or $^2\Pi$ electronic state.
These systems are ideal to analyze the interplay between the spin-orbit coupling and the 
rotational structure in the presence of external fields.
Furthermore, the Zeeman and Stark
effects might be comparable for moderate field strengths.
At the end of the 1990s, polar $^2\Sigma$ molecules were investigated in
congruent fields with the focus on their directional properties and  the possibility of trapping them~\cite{A908876H}. 
Recently, open shell  diatomic molecules, in $^2\Pi$ electronic states, 
exposed to combined electric and magnetic fields 
have been also analyzed from both theoretical and experimental point 
of views~\cite{PhysRevA.78.033433,PhysRevA.85.033427}. 
The avoided crossings in the field-dressed spectrum of OH were used to
transfer  population  between two states of opposite parity,
and the trap dynamics was observed in combined fields~\cite{PhysRevA.85.033427}. 
The theoretical analysis has been carried out under the assumption that the total angular momentum  
remains approximately constant, and, therefore, only the coupling between different $\Lambda$-doublet states has been 
taken into account~\cite{PhysRevA.78.033433,PhysRevA.85.033427}. 

The present study goes beyond this approximation and aims at an extended approach 
to the rotational motion of an open shell dimer in a $^2\Pi$ electronic state in combined
electric and magnetic fields. 
In particular, we describe it  within the rigid rotor approximation including  the 
fine structure interactions and the $\Lambda$-doubling effects.
Taking as prototype examples the LiO and OH radicals in their $^2\Pi$ electronic ground state,
we explore a wide range of field strengths and 
 two different regimes characterized by:
i) the field-dressed dynamics taking place within a certain rotational manifold;
and 
ii) the possibility of mixing states in neighbouring rotational manifolds  of the $^2\Pi_{3/2}$ fine structure 
component.
Our focus is on the energy shifts, the directional properties, and the hybridization
of the angular motion as either the electric or magnetic field strengths, or
the inclination angle between them is varied. 
We also investigate field-induced couplings between levels of the fine structure components
$^2\Pi_{1/2}$ and $^2\Pi_{3/2}$ by using the external fields. 
For a molecule with small rotational splitting, we show that the coupling between the spin and rotational degrees of freedom
could be achieved at realistic field strengths. It has been shown recently, that the
  spin-rotational coupling can be exploited for tailoring the interactions between ultracold
  molecules~\cite{micheli:nat_phys_2006,brennen:njp2007,lahaye:rep_prog_phys2099}. We show that
  similarly the presence of quasi-degenerate $\Lambda$-doublets of states provide versatile tools
  for shaping the properties of ultracold molecules.

The paper is organized as follows. In \autoref{sec:hamiltonian_symmetry}
we define our working Hamiltonian.  
In  \autoref{sec:results}, we discuss the numerical results for
two diatomic molecules with a $^2\Pi$ electronic ground state, OH and LiO, as
the field parameters are modified. 
In particular, we explore
three different cases: (i) for fixed magnetic field and four inclination
angles, we vary the electric field strength; (ii)
for fixed electric field and two inclination angles, the
magnetic field is increased; (iii) for fixed magnetic
and electric field strengths, the angle between them is
continuously changed from $0\degree$ to $90\degree$. Finally, 
for a model system, we investigate the possibility of mixing states 
from two different fine structure components. 
The conclusions  are provided in~\autoref{sec:conclusions}.

\section{The Hamiltonian of a linear rigid rotor in external fields}
\label{sec:hamiltonian_symmetry}
\begin{figure}
\begin{center}
\includegraphics[width=\columnwidth]{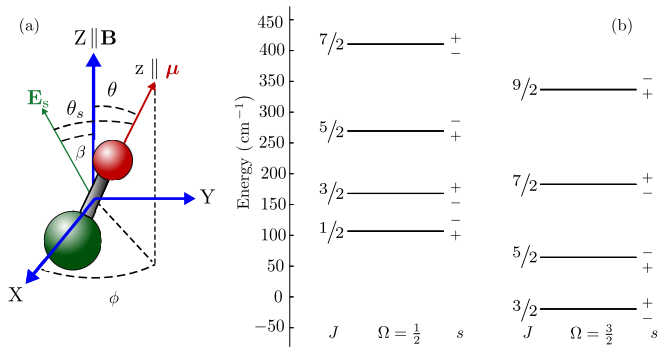}%
\caption{(a) Laboratory fixed frame, Euler angles,
schematic field configuration and diatomic molecule. 
(b) Field-free level structure of the OH molecule. 
The energy splitting due to the $\Lambda$-doubling is not visible on the scale of this figure, 
the plus and minus signs indicate if the even or odd component of a doublet is energetically higher or lower.
}%
\label{fig:fig_1}
\end{center}
\end{figure}

We consider a polar linear molecule in a $^2\Pi$ electronic state exposed  to a homogeneous
static electric field and a homogeneous static magnetic field. 
The field configuration is illustrated in \autoref{fig:fig_1}:
The magnetic field vector ${\bf B}$ points along the $Z$-axis of 
the laboratory fixed frame (LFF) $(X,Y,Z)$,
and the dc field $\bf{E}_{\textup{s}}$ is contained in the $XZ$-plane
forming an angle $\beta$ with the $Z$-axis.
The $z$-axis of the molecule fixed frame (MFF)
$(x,y,z)$ is defined by the permanent dipole moment of the
molecule  $\bm{\mu}$. These two frames are related by the Euler angles
$\theta$ and $\phi$, cf. \autoref{fig:fig_1}~(a). 
The system is described within the rigid rotor approximation,
assuming that the vibrational and electronic dynamics are not affected by the fields.
We have included the fine structure interactions and the $\Lambda$-doubling effects, 
this last term  being the largest second order correction for a $\Pi$-state~\cite{Brown2003}.
The effects due to higher order relativistic and hyperfine interactions have been neglected. 
Thus, the effective Hamiltonian~\cite{Brown2003} reads 
\begin{equation}
\label{eq:full_H}
  H=H_0+ H\ind{s} + H\ind{z}
  \end{equation}
where $H_0$  is the field-free Hamiltonian, and $H\ind{s}$  and 
$H\ind{z}$ stand for the interaction with the static electric and magnetic fields, respectively.
The field-free Hamiltonian is given by 
\begin{equation}
\label{eq:field_free_H}
 H_0 = H\ind{rot} + H\ind{so}  + H\ind{sr} + H_{\Lambda d} 
\end{equation}
where $H\ind{rot}$ is the rotational Hamiltonian~\cite{Brown2003}
\begin{equation}
\label{eq:rot_H}
H\ind{rot} ={\cal B}{\bf N}^2={\cal B}({\bf{J}}-{\bf{S}})^2
\end{equation}
with $\bf{N}$ being  the total angular momentum operator excluding  spin,  
$\bf{J}$ the total angular momentum operator excluding nuclear spin, $\bf S$ the electronic spin, 
and ${\cal B}$ the
rotational constant.
The spin-orbit coupling term reads 
\begin{equation}
\label{eq:so_H}
H\ind{so} =A\, T^1_0({\bf L})  T^1_0({\bf S})
\end{equation}
where $\bf L$ is the electronic orbital angular momentum and 
$A$ the spin-orbit constant. Note that we are using the spherical tensor notation~\cite{Zare1988}. 
The contribution due to the electronic spin-rotation coupling is given by
\begin{equation}
\label{eq:srot_H}
H\ind{sr} =\gamma\, T^1({\bf N})\cdot T^1({\bf S})
\end{equation}
where $\gamma$ is the spin-rotation constant.
For a  $^2\Pi$ electronic ground state,
the  $\Lambda$-doubling term reads
\begin{equation}
\label{eq:l_doubling_H}
H_{\Lambda d} =\sum_{m=\pm 1} e^{-2 i m\phi}
\Bigl(p T^2_{2m}({\bf S},{\bf N})-q T^2_{2m}({\bf N},{\bf N})
\Bigr)
\end{equation}
where $p$ and $q$ are the  $\Lambda$-doubling parameters.
The term $H_{\Lambda d}$ is 
a second order term causing a splitting between levels with different electronic 
angular momentum projection quantum numbers $\Lambda$.
It is due to the mixing of rotational states with even and odd parity, and the corresponding ones in the 
$\Sigma$ electronic state~\cite{Brown2003}.

The electric dipole moment couples to the static electric field, resulting in 
\begin{equation}
  \label{eq:hs}
  H_{\textup{s}}=-{\bm{\mu}}\cdot{\bf{E}}_{\textup{s}} 
=-\mu \Estatabs\cos\theta_{\textup{s}}
\end{equation}
with 
${\bf{E}}_{\textup{s}}=\Estatabs(\sin\beta \hat{X}+\cos\beta \hat{Z})$, 
and $\Estatabs$ being the electric  field strength.
The angle between the dipole moment ${\bm{\mu}}$ and this field is 
$\theta_{\textup{s}}$, cf. \autoref{fig:fig_1}, and 
$\cos\theta_{\textup{s}}=\cos\beta\cos\theta+\sin\beta\sin\theta\cos\phi$ 
with $0\degree\le\beta\le180\degree$. 

The interaction with the magnetic field is given by 
\begin{equation}
  \label{eq:hz}
  H\ind{z}=\mu_B {\bf B} \cdot (g_L{ \bf L} + g_s {\bf S})
\end{equation}
where $\mu_B=e\hbar/2m$ is the Bohr magneton, and $g_L$ and $g_s$, are the electron 
orbital and spin gyromagnetic ratios, respectively, which can be approximated by
 $g_L\approx1$ and $g_s\approx2$. 

Here, we consider molecules having a spin-orbit constant $A$ larger than the rotational constant
${\cal B}$.
For the description of these systems, 
the Hund's  case (a) coupling is suited best~\cite{Brown2003}. 
The basis set is formed by eigenstates of the commuting operators 
{$L_z$, $\bf{ S}^2$, $S_z$, ${\bf{J}}^2$, $J_Z$, and $J_z$}. 
The operators  $L_z$,  $J_z$ and  $S_z$ are the projections of  the electronic orbital angular momentum
$\bf L$,  total angular momentum $\bf J$ and spin $\bf S$
on the $z$-axis of the MFF, respectively, whereas  $J_Z$ is the projection of  $\bf J$ 
on the LFF $Z$-axis. 
Since this study is restricted to the vibrational ground state of the electronic
ground state, the electronic and vibrational Hamiltonians have not been included.
The eigenstates of this basis $\ket{\Lambda S\Sigma J M_J  \Omega}$  are labeled by    
$ {\Lambda, S, \Sigma, J,  M_J, \Omega}$,
with $\Omega=\Lambda+\Sigma$. 
For a $^2\Pi$ electronic  state, $S=1/2$ and  $\Lambda=\pm1$. 
Performing a transformation to a basis of parity eigenstates
\begin{eqnarray}
\ket{J  M_J  \Omega s} 
&= &\frac{1}{\sqrt{2}}
\Bigl(\ket{\Lambda S\Sigma J M_J \Omega} \nonumber \\ 
&+&
s (-1)^{J-S} \ket{-\!\Lambda\, S -\!\Sigma\,  J  M_J -\!\Omega}\Bigr)
\label{eq:parity_states}
\end{eqnarray}
the label $\Lambda$ becomes obsolete and is replaced by the parity $s=\pm1$. 
Thus,  only  $\Omega$, $J$, $M_J$, and $s$ are needed
to uniquely label the states.

The field-free Hamiltonian is invariant under any rotation, and $J$, $M_J$ and the parity $s$,  are
good quantum numbers, whereas, $\Omega$ is not well defined.
The field-free states having different $\Omega$ are coupled, and their labeling is  
based on the adiabatic limit of vanishing spin-rotational coupling. There are two manifolds of fine  
structure levels 
$^2\Pi_{3/2}$ and $^2\Pi_{1/2}$, each one 
consisting of several rotational levels (cf. Fig. \ref{fig:fig_1}~(b)). 
For molecules having a negative fine structure constant, $A<0$, 
e.g., the OH, LiO and NaO molecules, the ground state  has $\Omega=3/2$. 
The $\Lambda$-doubling splitting between states  of  different parity is small compared to the 
rotational splitting. The field-free states are degenerate in $M_J$.

The symmetries of this system are significantly reduced when the fields are applied.
In the presence of a static electric field, the Hamiltonian is invariant under arbitrary rotations 
around the field axis ${\cal C}_{\bf{E}_s}(\delta)$ and reflections in any
plane containing the field axis, being $M_J$ a good quantum number if $\beta=0\degree$.
If only the magnetic field is applied, the symmetries of the Hamiltonian consist on 
arbitrary rotations around the LFF $Z$-axis ${\cal C}_Z(\delta)$
and inversions with respect to the origin of coordinates. In this case,  $M_J$
and $s$ remain as good quantum numbers. 
For parallel or antiparallel fields, the Hamiltonian is invariant under any arbitrary
rotations around the field axis ${\cal C}_Z(\delta)$ and $M_J$ is a good quantum number. 
If the fields are perpendicular the reflection in the plane perpendicular 
to the magnetic field, $\sigma_{XY}$ is a symmetry. For any other angle, \ie,
 $\beta\ne 0\degree, 90\degree$ or $180\degree$, 
all the symmetries of the Hamiltonian are broken.

The time-independent Schr\"odinger equation associated to the
Hamiltonian \eqref{eq:full_H} is solved by a basis set expansion in terms of 
the functions \eqref{eq:parity_states}. For reasons of addressability,  
we will label the field-dressed states 
as $\ket{J  M_J  \Omega s}$, even if $J$, $M_J$ or $s$ are not good quantum numbers. 
Thus, $\ket{J  M_J  \Omega s}$ refers to the level that is adiabatically
connected as $\Estatabs$, $\Magabs$, and/or
$\beta$ are modified with the field-free state  $\ket{J  M_J  \Omega s}$.

\section{Results}
\label{sec:results}
The molecules LiO, NaO, OH and NO have   $^2\Pi$ electronic ground states. 
The parameters of the effective Hamiltonian are 
listed in~\autoref{tab:molecular_constant}~\cite{Brown2003,yamada:jcp1989_vol91,yamada:jcp1989}.
For LiO ~\cite{yamada:jcp1989_vol91} and NaO~\cite{yamada:jcp1989}, the values of the spin-rotation
constants are not available in the literature to the best of our knowledge, so we have estimated them
  using an approximation based on a pure precession hypothesis 
$\gamma\approx -p/2$~\cite{jms:brown_1979,jms:zare_1973,jms_74_488:brown_1979,Lefebvre1986}.
In this work, we consider two of these molecules,   LiO and OH,  as benchmarks to illustrate our results. 
While both molecules have spin-orbit constants of the same order of magnitude, 
the  rotational constant of OH is around $15.4$ times larger than in LiO.
For OH, there are two rotational manifolds of $^2\Pi_{3/2}$
energetically below the first rotational manifold with $J=1/2$ of $^2\Pi_{1/2}$, see
 the states with $J=3/2$ and $5/2$ of $^2\Pi_{3/2}$ 
in~\autoref{fig:fig_1}~(b). Opposite to this, for LiO, $8$ rotational manifolds with 
$\Omega=3/2$ have energies smaller than the rotational ground state of $^2\Pi_{1/2}$. 
For both systems, the energy gap to the neighbouring $\Sigma$ electronic state is large enough, so that its influence can be described
by the $\Lambda$-doubling~\cite{Brown2003}.
  The  contribution of the hyperfine structure is negligible, being the contribution of the quadrupole
  moment for the LiO in the order of tenths of MHz~\cite{Brown2003,freund:jcp_1972,yoshimine:jcp1972}. 
Compared to OH, LiO has a larger dipole moment and a smaller rotational constant, thus,  the same   electric 
field strength would provoke a larger impact on the LiO rotational dynamics.

\renewcommand{\thefootnote}{\thempfootnote}
\begin{table}[h]
  \caption{Parameters of the effective Hamiltonian for the $^2\Pi$ ground state of the OH, LiO, NaO and
    NO molecules. 
  }
{\begin{tabular}{ccccc}
\toprule
Molecule &
OH	 & LiO\footnote{\label{aaa}The spin-rotation constants of  LiO~\cite{yamada:jcp1989_vol91}
and NaO~\cite{yamada:jcp1989} molecules has been estimated using the approximation $\gamma\approx
-p/2$~\cite{jms:brown_1979,jms:zare_1973,jms_74_488:brown_1979}.}\label{tab:molecular_constant}
 & NaO\footnotemark[\value{mpfootnote}] & NO  \\
\hline
${\cal B}$  (cm$^{-1}$)  & $18.535$  & $  1.204$& $  0.422 $  &    $1.696$ \\
 $A$ (cm$^{-1}$)   & $-139.051$ & $  -111.672$  & $  -107.151 $ &  $123.15$ \\
$\gamma$ (cm$^{-1}$)  &$-0.119$ & $ -0.105$ &   $  -4.42\times 10^{-2}$  &$-6.47\times 10^{-3}$\\
  $p$  (cm$^{-1}$)    &$0.235$    & $  0.210  $ &$  8.84\times 10^{-2}$ & $1.17\times 10^{-2}$ \\
$q$ (cm$^{-1}$)  &$-0.039$  &  $  -1.89\times 10^{-3}$ & $  6.23\times 10^{-4}$&$  9.41\times 10^{-5}$ \\
 $\mu$ (D)&$1.655$&   $6.5$  &$8.7$ & $ 0.159$\\ 
\botrule
  \end{tabular}}
\end{table}

\subsection{Influence of the electric-field strength}

\begin{figure}
\begin{center}
\includegraphics[width=\columnwidth]{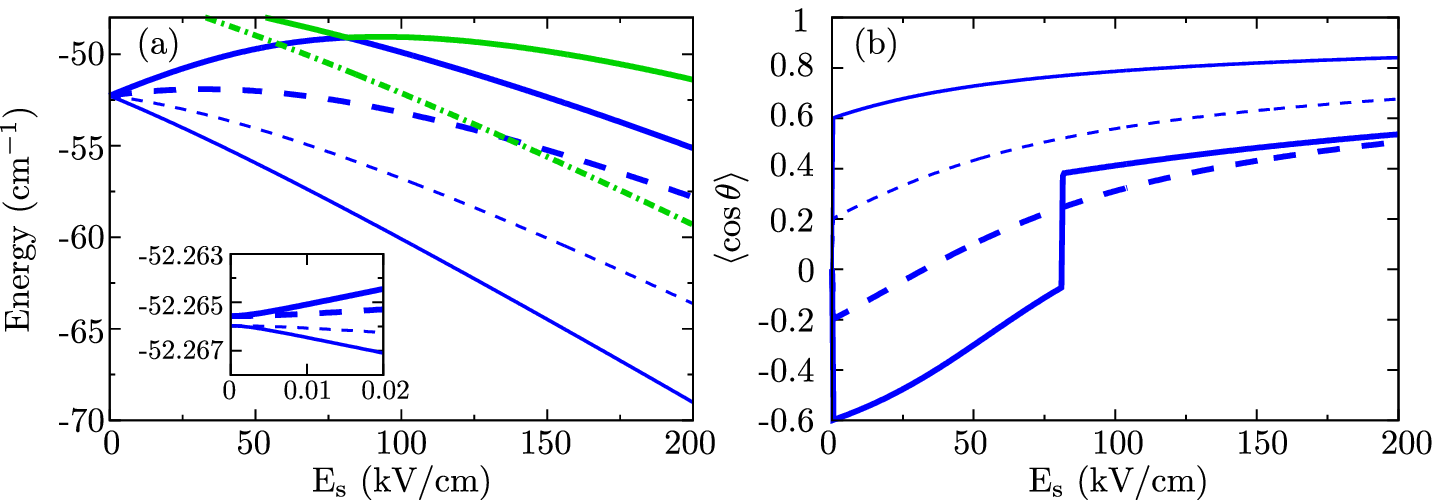}%
\caption{For LiO, we present (a) the energy and (b) the expectation value $\expected{\cos\theta_{\textup{s}}}$ 
versus the electric field strength
of the states
$\ket{\nicefrac{3}{2}, \nicefrac{3}{2}, \nicefrac{3}{2},\pm1}$ (blue thick and thin solid), 
and 
$\ket{\nicefrac{3}{2}, \nicefrac{1}{2}, \nicefrac{3}{2},\pm 1}$ (blue thick and thin  dashed).
Some states with $J=5/2$ are also plotted:
$\ket{\nicefrac{5}{2}, \nicefrac{5}{2}, \nicefrac{3}{2},1}$ (green thick dot-dashed)
and
$\ket{\nicefrac{5}{2}, \nicefrac{3}{2}, \nicefrac{3}{2},1}$ (green thick solid). 
Due to the degeneracy in $|M_J|$, only those states with $M_J>0$ are presented.
The  inset in panel (a) shows the energy splitting for weak electric field.
The field configuration is $\beta=0\degree$ and $\Magabs= 0$~T. }
\label{fig:fig_static_field_only}
\end{center}
\end{figure}

\begin{figure}
\begin{center}
\includegraphics[width=\columnwidth]{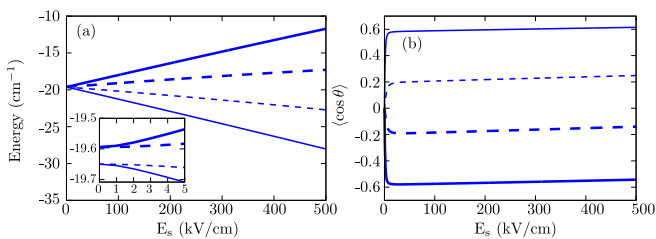}%
\caption{For OH, 
we present  (a) the energy and (b) the expectation value $\expected{\cos\theta_{\textup{s}}}$ versus the electric field 
strength. 
The inset in panel (a) shows the energy splitting for weak electric field. 
The labeling of the states is done as in \autoref{fig:fig_static_field_only}, and 
due to the degeneracy in $|M_J|$, only those states with $M_J>0$ are plotted. 
The field configuration is $\beta=0\degree$ and $\Magabs= 0$~T.}
\label{fig:fig_static_field_only_OH}
\end{center}
\end{figure}

We start by analyzing the impact of a static electric field taking $\beta=0\degree$. 
We restrict this study to the lowest lying eight states with  $\Omega=3/2$ and $J=3/2$. 
Note that they well represent the main physical features observed in the overall molecular
dynamics, and similar behavior and properties are, therefore, obtained for highly excited levels.  
For the  $\Omega=3/2$ and $J=3/2$ states of LiO,
\autoref{fig:fig_static_field_only}~(a) and \autoref{fig:fig_static_field_only}~(b) show 
 the energy and $\expected{\cos\theta}$,
respectively, as a function of $\Estatabs$, for $\beta=0\degree$ and
$\Magabs= 0$~T. 
The corresponding results for OH are presented in 
\autoref{fig:fig_static_field_only_OH}~(a) and \autoref{fig:fig_static_field_only_OH}~(b). 

The Stark interaction \eqref{eq:hs} couples states with opposite parity, if $\beta=0\degree$,  
$M_J$ is a good quantum number and the levels with $\pm M_J$ are still degenerate, 
although the field-free degeneracy in  $M_J$ is broken. 
In the weak dc-field regime, the two  $\Lambda$-doubling components are strongly coupled because
they are energetically close. In addition, the coupling to the nearest rotational levels is rather
weak because they are far apart in the spectrum.
The $\Lambda$-doubling splitting within the $\Omega=3/2$ and $J=3/2$ manifold are 
$\sim 10^{-4}$~cm$^{-1}$ and $\sim 5.6\times 10^{-2}$~cm$^{-1}$ for  LiO and OH, respectively; 
whereas they are separated by $5.96$~cm$^{-1}$ and $44.5$~cm$^{-1}$ to the levels with
$\Omega=3/2$ and $J=5/2$. 
Thus, in the weak dc-field regime, the system could be described as
a two state model formed by the even and odd parity levels with the same $M_J$.  
Only for very strong static electric fields, the couplings to the states within the next
rotational manifold might become more important. 
Within this approximation, the Stark effect correction
reads   
\begin{equation}
  \label{eq:hz_spliting}
  \Delta E=\pm \sqrt{\frac{E_\Lambda^2}{4}+\mu_{eff}^2\Estatabs^2}
\end{equation}
for the $s=\pm1$ states with $E_\Lambda$ being the $\Lambda$-doublet splitting, 
and  $\mu_{eff}=-\mu |M_J|\Omega/J(J+1)$. 
The odd and even parity states are initially high- and low-field  seekers, respectively. 
If the $\Lambda$-doublet splitting is small, 
even a very weak electric field might induce an efficient orientation, 
but the two states with different parity have their effective electric
dipole moments pointing in opposite directions. 
For LiO and $\Estatabs= 20$~V/cm, 
$\expected{\cos\theta}=\pm0.60$ and $\pm0.20$ for the even and odd 
states with $|M_J|=3/2$ and $1/2$, respectively. 
For OH, a stronger dc-field is needed to achieve a similar orientation, \eg, 
$\expected{\cos\theta}=\pm0.57$ and
$\pm0.19$ for the even and odd states with
$|M_J|=3/2$ and $1/2$, respectively, for  $\Estatabs= 10$~kV/cm.

By further increasing $\Estatabs$, the couplings to the next rotational manifold, and, therefore,
the contribution of these 
states  to the field-dressed dynamics should become more important.
However, this is not the case for OH,  and these couplings are still  small even 
for strong electric fields due to its large rotational constant. 
For all the $J=3/2$ levels, the variation of $\expected{{\bf{J}}^2}$ 
compared to the field-free value  is below $0.8\%$ even 
for $\Estatabs= 500$~kV/cm.
Then, the OH states keep their low- and high-field seeking character 
and their orientations vary smoothly, cf. \autoref{fig:fig_static_field_only_OH}. 

\begin{figure}
\begin{center}
\includegraphics[width=\columnwidth]{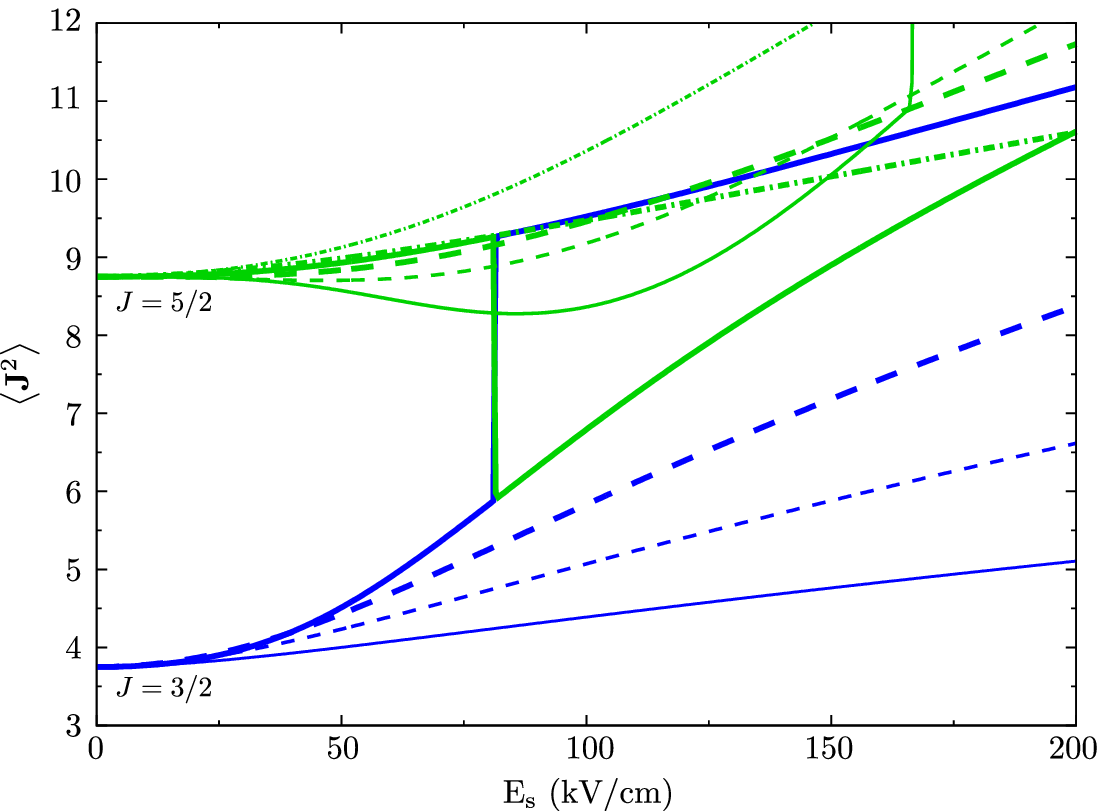}%
\caption{For LiO, we show the expectation value $\expected{{\bf{J}}^2}$ versus the electric field strength 
of the states
$\ket{\nicefrac{3}{2}, \nicefrac{3}{2}, \nicefrac{3}{2},\pm1}$  (blue thick and thin solid), 
$\ket{\nicefrac{3}{2}, \nicefrac{1}{2}, \nicefrac{3}{2},\pm 1}$ (blue thick and thin dashed),
$\ket{\nicefrac{5}{2}, \nicefrac{5}{2}, \nicefrac{3}{2},\pm 1}$  (green thick and thin  dot-short-dashed), 
$\ket{\nicefrac{5}{2}, \nicefrac{3}{2}, \nicefrac{3}{2},\pm 1}$ (green thick and thin  solid),
and 
$\ket{\nicefrac{5}{2}, \nicefrac{1}{2}, \nicefrac{3}{2},\pm 1}$ (green thick and thin  dashed).
Due to the degeneracy in $|M_J|$, only those states with $M_J>0$ are plotted.
The field configuration is $\beta=0\degree$ and $\Magabs= 0$~T. 
}%
\label{fig:fig_static_field_only_j2_LiO}
\end{center}
\end{figure}

In contrast, these couplings to high $J$-value states leave their fingerprints in the LiO spectrum 
even at moderate electric field strength. 
To illustrate the hybridization of the total  angular momentum in LiO,
we have plotted in~\autoref{fig:fig_static_field_only_j2_LiO}
$\expected{{\bf{J}}^2}$ versus $\Estatabs$ for the 
$J=3/2$ and $J=5/2$ levels of $^2\Pi_{3/2}$.
For weak electric fields,  $\expected{{\bf{J}}^2}$ presents   a plateau-like behaviour around
the field-free values $\expected{{\bf{J}}^2}=3.75$ and  $8.75$ for the $J=3/2$ and $J=5/2$ states, respectively.
For the $\ket{\nicefrac{5}{2}, \nicefrac{\pm 3}{2},\nicefrac{3}{2}, -1}$  and $\ket{\nicefrac{5}{2}, \nicefrac{\pm 1}{2},\nicefrac{3}{2}, -1}$ levels, the coupling to states in the lowest 
rotational manifold is initially dominant, and  $\expected{{\bf{J}}^2}$  decreases as $\Estatabs$  is increased,
reaches a broad minimum increasing thereafter.
For the rest,  $\expected{{\bf{J}}^2}$ monotonically increases due to the couplings to states with  higher field-free 
$J$-values. 
The $\ket{\nicefrac{3}{2}, \nicefrac{\pm3}{2},\nicefrac{3}{2}, 1}$ levels suffer an avoided crossing with 
$\ket{\nicefrac{5}{2}, \nicefrac{\pm3}{2}, \nicefrac{3}{2}, 1}$ from the next rotational manifold 
for
$\Estatabs\approx 81.437$~kV/cm
of width $\Delta E=1.01\times 10^{-3}$~cm$^{-1}$, and
they interchange their intrinsic character. 
Then, the $\ket{ \nicefrac{3}{2}, \nicefrac{\pm3}{2},\nicefrac{3}{2}, 1}$ levels become oriented and
acquire a high-field seeking character, see~\autoref{fig:fig_static_field_only}. 
In the strong dc-field regime, all the $J=3/2$ states are high-field seekers, and 
their orientation monotonically increases as $\Estatabs$ is increased. 
For $\Estatabs= 200$~kV/cm, the levels  with 
$|M_J|=3/2$ and $1/2$ and $s=1$ have achieved a significant orientation with 
$\expected{\cos\theta}=0.84$ and $0.68$, respectively, 
whereas $\expected{\cos\theta}=0.54$ and $0.51$ for the corresponding states with odd parity.
They also show a strong hybridization of the angular motion, \eg, 
for $\Estatabs= 200$~kV/cm,  $\expected{{\bf{J}}^2}=5.11$ and 
$6.62$ ($8.37$ and $11.18$) for the odd (even) $|M_J|=3/2$ and $1/2$ states. 

\subsection{Influence of the magnetic field strength}
\begin{figure}
\begin{center}
\includegraphics[width=\columnwidth]{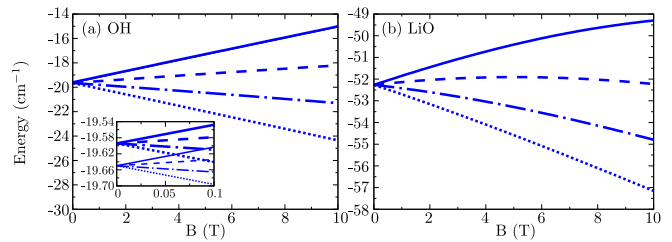}%
\caption{For the $\Omega=3/2$ and $J=3/2$ states,
we show their energy  versus the magnetic field strength for (a) OH and (b) LiO. 
The inset in panel (a) shows the energy splitting for  weak  magnetic field. 
The states are 
$\ket{\nicefrac{3}{2},  \nicefrac{3}{2}, \nicefrac{3}{2},\pm 1}$ (thick and thin solid),
$\ket{\nicefrac{3}{2}, \nicefrac{-3}{2}, \nicefrac{3}{2},\pm1}$ (thick and thin dotted),
$\ket{\nicefrac{3}{2},  \nicefrac{1}{2}, \nicefrac{3}{2},\pm 1}$ (thick and thin dashed),
and 
$\ket{\nicefrac{3}{2},  \nicefrac{-1}{2}, \nicefrac{3}{2},\pm 1}$ (thick and thin  dot-dashed). 
Due to the small $\Lambda$-doubling the two states forming
 a $\Lambda$-doublet cannot be distinguished on the scale of this figure. 
The field configuration is $\beta=0\degree$ and $\Estatabs= 0$~V/cm.}%
\label{fig:fig_magnetic_field_olny}
\end{center}
\end{figure}

Here, we investigate the evolution of the energy as the magnetic field strength is increased.
For the OH and LiO molecules,
\autoref{fig:fig_magnetic_field_olny}~(a) and \autoref{fig:fig_magnetic_field_olny}~(b) show 
the energy of the $\Omega=3/2$ and $J=3/2$ states  as a function of $\Magabs$
with $\Estatabs= 0$~V/cm and $\beta=0\degree$. 
In this field configuration, $M_J$ and $s$ are good quantum numbers.  
The  states with $M_J<0$ $(>0)$ are high- (low-) field seekers as $\Magabs$ is varied, 
and the two components of a $\Lambda$-doublet run parallel. 
The OH molecule presents the linear Zeeman effect, \ie, the energy linearly depends on the magnetic
field strength $\Magabs$, and the states with
different parity and $M_J$ suffer exact crossings, see inset in \autoref{fig:fig_magnetic_field_olny}, when $\Magabs$ is increased. 
For OH, the mixing with  states of the neighbouring rotational manifold is negligible even at  
$\Magabs= 10$~T, where 
the variation of $\expected{{\bf{J}}^2}$ compared to its field-free value is below $1.2\%$, and the alignment is also 
weakly affected.

\begin{figure}
\begin{center}
\includegraphics[width=\columnwidth]{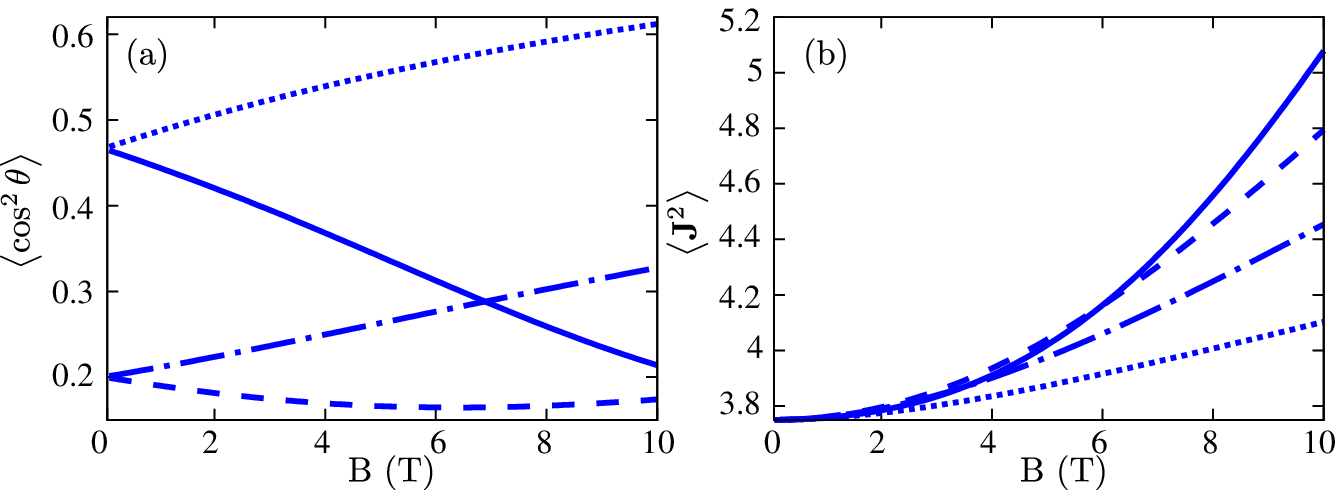}%
\caption{For the $\Omega=3/2$ and $J=3/2$ states of LiO,
we show their 
expectation values (a) $\expected{\cos^2\theta_{\textup{s}}}$  and 
(b) $\expected{{\bf{J}}^2}$ 
 versus the magnetic field strength. 
The field configuration is $\beta=0\degree$ and $\Estatabs= 0$~V/cm.
The labeling of the states is done as in \autoref{fig:fig_magnetic_field_olny}. 
}%
\label{fig:fig_magnetic_field_olny_j2_cos2}
\end{center}
\end{figure}

Due to the small $\Lambda$-doubling splitting of LiO, the states with even and odd parity are very
close in energy and indistinguishable on the scale 
of \autoref{fig:fig_magnetic_field_olny}~(b). If the magnetic field is strong enough, the coupling to 
levels of the neighbouring $J$-manifold becomes more important, and the linear Zeeman behaviour is lost.
The alignment $\expected{\cos^2\theta}$ and hybridization of the angular motion
$\expected{{\bf{J}}^2}$ are presented in~\autoref{fig:fig_magnetic_field_olny_j2_cos2} (a) and (b), respectively. 
The alignment of the states with $M_J<0$ $(>0)$ increases (decreases) as $\Magabs$ is increased.
Only the levels with $M_J=-3/2$
present moderate alignment with $\expected{\cos^2\theta}>0.5$ for $\Magabs\gtrsim 1.7$~T. 
As in the case of a static electric field, 
$\expected{{\bf{J}}^2}$ monotonically increases with $\Magabs$.

\subsection{Influence of combined magnetic and electric fields}

\begin{figure}
\begin{center}
\includegraphics[width=\columnwidth]{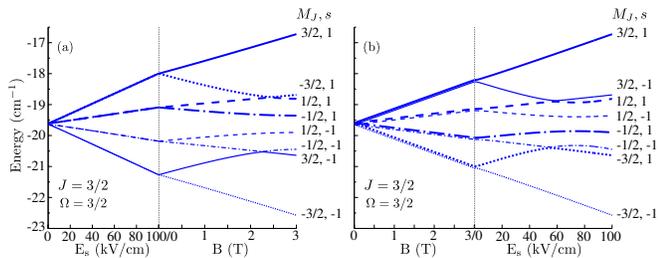}%
\caption{The $J=3/2$ manifold of $^2\Pi_{ \nicefrac{3}{2}}$ is presented at intermediate field strengths. 
The relative angle between the fields is   $\beta=30\degree$. 
The labels of the states are put according to the  adiabatic following.} 
\label{fig:states_labeling}
\end{center}
\end{figure}

The rotational dynamics is  drastically modified when the molecules are exposed to
combined magnetic and electric fields. 
 The Zeeman and Stark interactions break different symmetries
of the field-free Hamiltonian, and  the order in which the fields are turned on determines the evolution
of the field-dressed states. Indeed, their labels depend on the path followed on the
parameter space, $\Estatabs$,  $\Magabs$ and $\beta$, to reach a certain field configuration.
This phenomenon, called monodromy,  
has been  previously observed in diatomic and polyatomic molecules in external 
fields~\cite{schleif:pra2007,kozin:jcp2003}. 
In \autoref{fig:states_labeling}, it is illustrated by the evolution
of the $\Omega=3/2$ and $J=3/2$ states of OH increasing first the electric field and
then the magnetic one (left panel) or vice versa (right panel) with  $\beta=30\degree$.

These two pathways through the parameter space 
lead to different state labels at the final point. 
The main reason is that, as indicated above, by modifying the field configuration
we change the symmetries of the 
system and on different parameter pathways the symmetries are broken in different order.
The energetic ordering
depends on which field is present at first.
If the electric field is turned on first, the parity in the field-free limit 
determines the low or high-field seeking character of a level, see~\autoref{fig:states_labeling}~(a),
whereas in the case of a magnetic field, it is the sign of $M_J$,
cf. \autoref{fig:states_labeling}~(b). 
It shall be noted that the electric field immediately mixes states of different parity. So the
  label $s$ is given according to adiabatic following and  no longer has the meaning of the
  parity of the eigenstate but indicates if a state is oriented ($s=-1$) or
  anitoriented ($s=1$). 
As the second field is switched on, the previous energetic ordering of the states is kept since
there are no more good quantum 
numbers and all crossings encountered in the spectrum are avoided, except for the case
$\beta=90\degree$.  
This non-uniqueness of the state labeling is typical for systems which cannot be described
using a single set of irreducible representations in the whole  parameter space.  

\subsubsection{Constant magnetic field strength and increasing electric field strength}

\begin{figure}
\begin{center}
\includegraphics[width=\columnwidth]{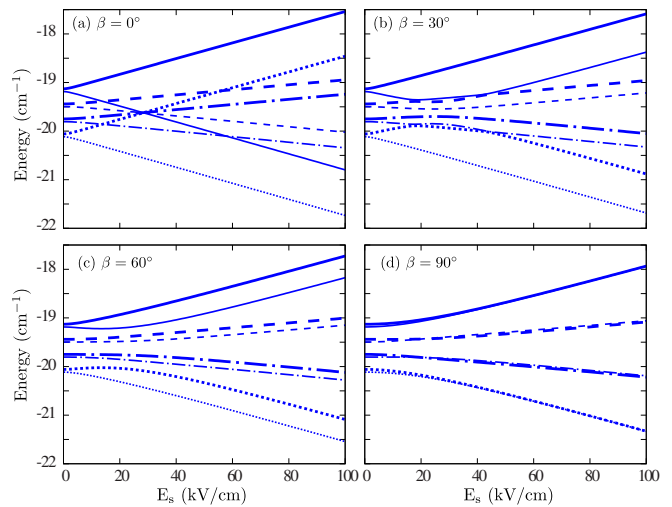}%
\caption{For the $\Omega=3/2$ and $J=3/2$ states of OH,
we present the energy  versus the electric field strength for 
 (a) $\beta=0\degree$,  (b) $\beta=30\degree$,  (b) $\beta=60\degree$ and
  (c) $\beta=90\degree$, with $\Magabs= 1$~T.
The labeling of the states is done as in~\autoref{fig:fig_magnetic_field_olny}. 
}%
\label{fig:fig_OH_B_fixed_E_versus_Es}
\end{center}
\end{figure}

\begin{figure}
\begin{center}
\includegraphics[width=\columnwidth]{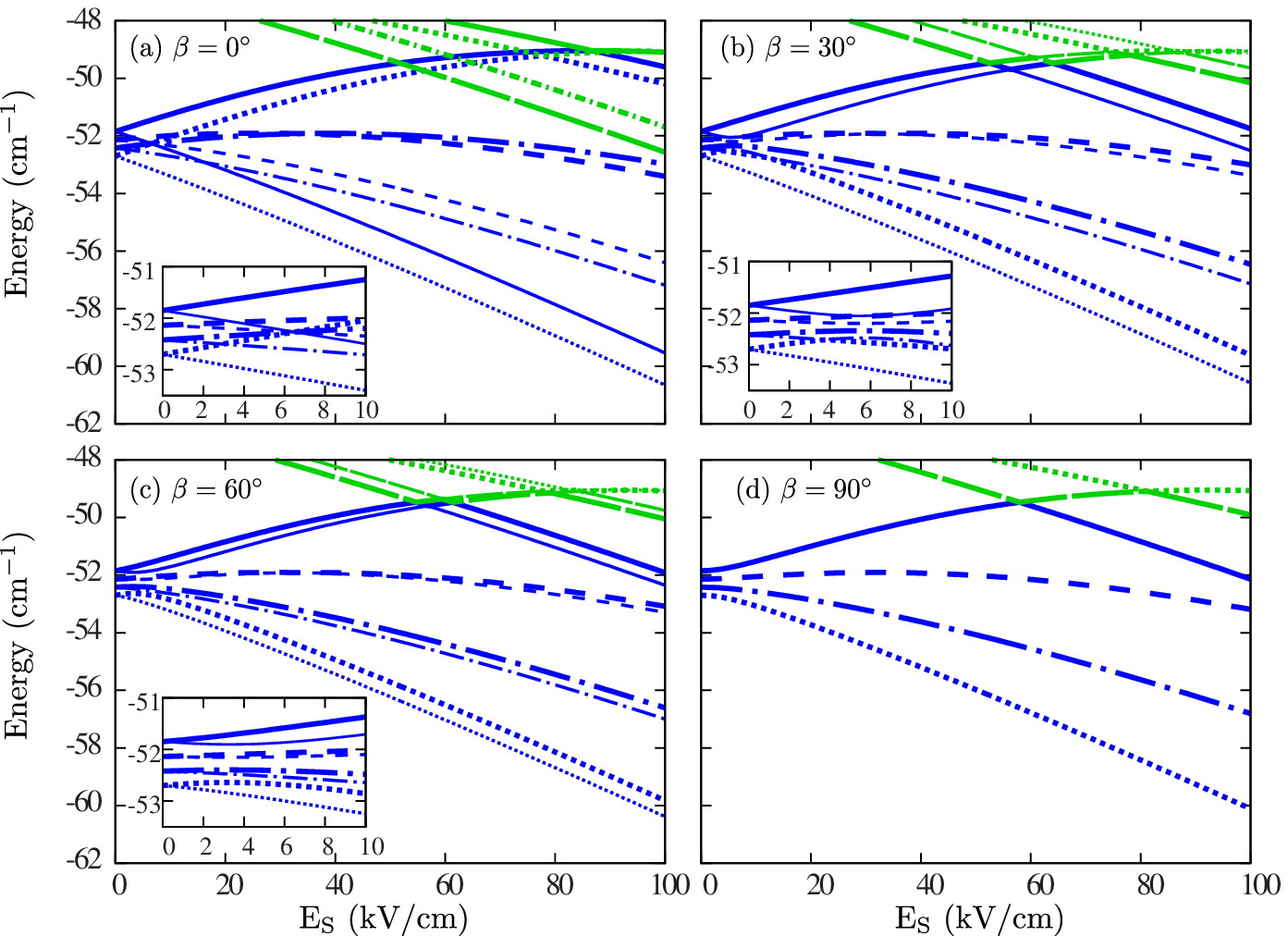}%
\caption{Same as \autoref{fig:fig_OH_B_fixed_E_versus_Es} but for LiO.
The insets show the weak electric field regime. 
Some states from the rotational manifold $J=5/2$ are included:
$\ket{\nicefrac{5}{2}, \nicefrac{5}{2}, \nicefrac{3}{2},1}$  (green thick  dot-short-dashed), 
$\ket{\nicefrac{5}{2}, \nicefrac{-5}{2}, \nicefrac{3}{2},\pm1}$  (green thick  and thin long-dashed), 
$\ket{\nicefrac{5}{2}, \nicefrac{3}{2}, \nicefrac{3}{2},1}$ (green thick solid),
and 
$\ket{\nicefrac{5}{2}, \nicefrac{-3}{2}, \nicefrac{3}{2},\pm1}$ (green thick and thin dotted).
}%
\label{fig:fig_LiO_B_fixed_E_versus_Es}
\end{center}
\end{figure}

\begin{figure}
\begin{center}
\includegraphics[width=\columnwidth]{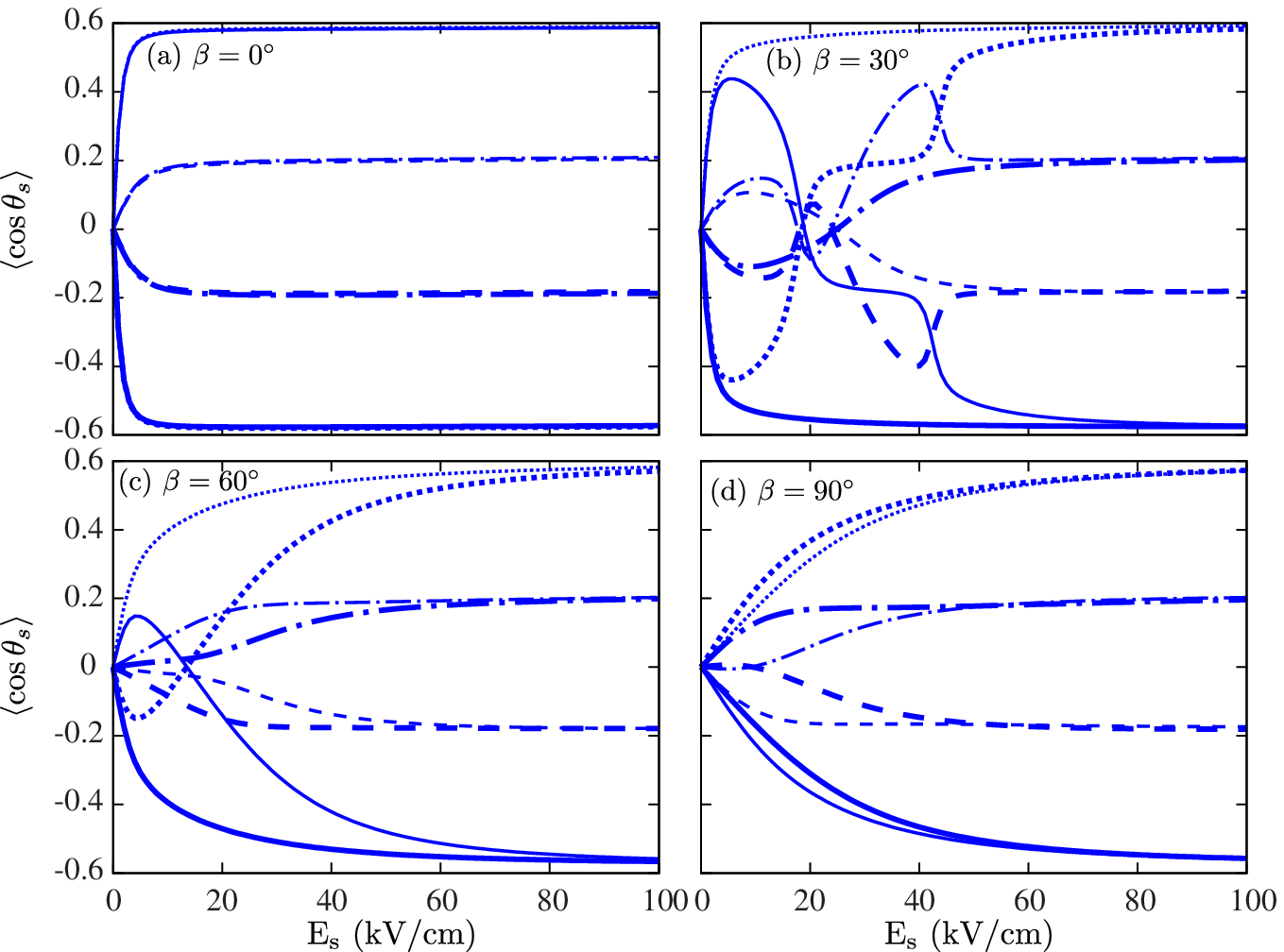}%
\caption{For the $\Omega=3/2$ and $J=3/2$ states of OH,
the expectation value $\expected{\cos\theta_{\textup{s}}}$ is shown versus the electric field strength for 
 (a) $\beta=0\degree$,  (b) $\beta=30\degree$,  (c) $\beta=60\degree$ and
  (d) $\beta=90\degree$, with $\Magabs= 1$~T.
The labeling of the states is done as in \autoref{fig:fig_magnetic_field_olny}. 
}%
\label{fig:fig_OH_B_fixed_cos_versus_Es}
\end{center}
\end{figure}

\begin{figure}
\begin{center}
\includegraphics[width=\columnwidth]{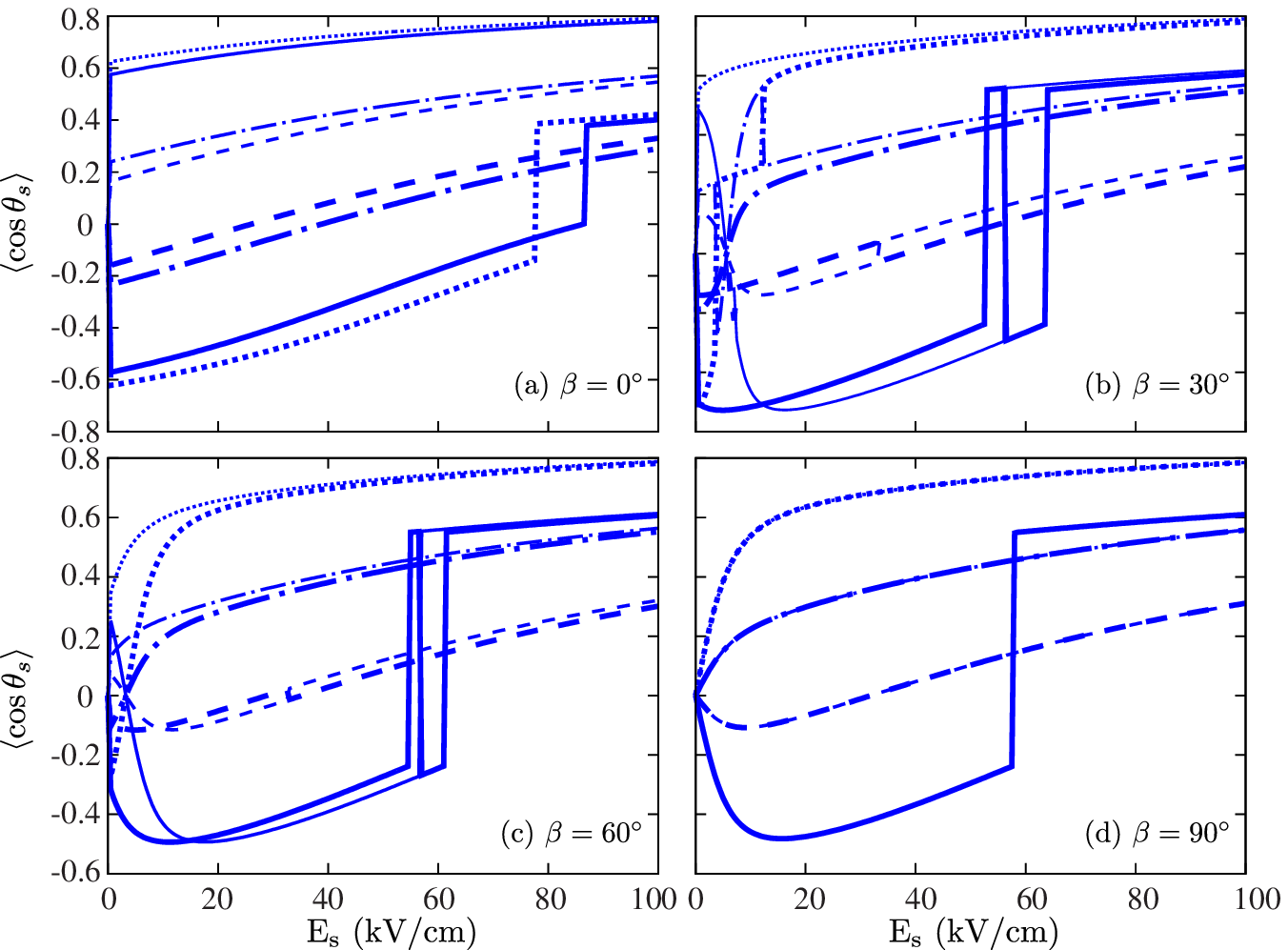}%
\caption{Same as \autoref{fig:fig_OH_B_fixed_cos_versus_Es}
but for LiO. }%
\label{fig:fig_LiO_B_fixed_cos_versus_Es}
\end{center}
\end{figure}

Now, we  consider a field configuration in which after turning on a magnetic field of 
$\Magabs= 1$~T, the electric field is switched on forming an angle
$\beta$ with the LFF $Z$-axis. For LiO and OH, 
the variation of the energies of the states with $\Omega=3/2$ and $J=3/2$ 
as a function of $\Estatabs$ is presented in 
\autoref{fig:fig_OH_B_fixed_E_versus_Es}~(a), (b), (c) and (d) and 
\autoref{fig:fig_LiO_B_fixed_E_versus_Es}~(a), (b), (c) and (d), 
for the inclination angles $\beta=0\degree$, $30\degree$, $60\degree$ and $90\degree$,
respectively. 
The corresponding results for the orientation cosines $\expected{\cos\theta_{\textup{s}}}$ 
are presented in~\autoref{fig:fig_OH_B_fixed_cos_versus_Es} and
\autoref{fig:fig_LiO_B_fixed_cos_versus_Es}.  


Let us start analyzing the results for the parallel-field configuration. 
In this case, $M_J$ is still a good quantum number, and within this manifold the states suffer exact crossings.
At $\Estatabs= 0$~V/cm and $\Magabs= 1$~T,
the $M_J$-degeneracy is already lifted and the $8$ states are identified on the OH spectrum, whereas for LiO, the 
$\Lambda$-doubling is so small that the even and odd parity states are still quasidegenerate. As $\Estatabs$ is increased
the odd (even) parity states become high- (low-) field seekers. 
The levels 
$\ket{ \nicefrac{3}{2}, \nicefrac{-3}{2}, \nicefrac{3}{2},1}$, 
$\ket{ \nicefrac{3}{2}, \nicefrac{-1}{2}, \nicefrac{3}{2},1}$, 
$\ket{ \nicefrac{3}{2}, \nicefrac{1}{2}, \nicefrac{3}{2},-1}$ and 
$\ket{ \nicefrac{3}{2}, \nicefrac{3}{2}, \nicefrac{3}{2},-1}$  
have the same energy for 
$\Estatabs\approx 28$~kV/cm and  $\Estatabs\approx 6.4$~kV/cm
for OH and LiO respectively.
For LiO, 
we find an avoided crossing among the levels
$\ket{ \nicefrac{3}{2}, \nicefrac{-3}{2}, \nicefrac{3}{2},1}$ 
and 
$\ket{ \nicefrac{5}{2}, \nicefrac{-3}{2}, \nicefrac{3}{2},1}$ 
at
$\Estatabs\approx 77.782$~kV/cm of width $\Delta E=8.5\times 10^{-4}$~cm$^{-1}$, 
a second one is encounter between  the states 
$\ket{ \nicefrac{3}{2}, \nicefrac{3}{2}, \nicefrac{3}{2},1}$ 
and 
$\ket{ \nicefrac{5}{2}, \nicefrac{3}{2}, \nicefrac{3}{2},1}$ 
at
$\Estatabs\approx  86.7827$~kV/cm of width $\Delta E=1.15\times 10^{-3}$~cm$^{-1}$.
For this system,  all the states are  high-field seekers in the strong field regime. 

For non-parallel fields $\beta\ne90\degree$, there are no symmetries left on the rigid-rotor Hamiltonian. 
The real crossings appearing for $\beta=0\degree$ become avoided, in particular, the states
$\ket{ \nicefrac{3}{2}, \nicefrac{-3}{2}, \nicefrac{3}{2},1}$ and 
$\ket{ \nicefrac{3}{2}, \nicefrac{-1}{2}, \nicefrac{3}{2},-1}$ 
suffer two consecutive avoided crossings at
$\Estatabs\approx  3.7835$~kV/cm 
with  
$\Delta E=2.2\times 10^{-4}$~cm$^{-1}$ 
and 
$\Estatabs\approx  12.350$~kV/cm
with  
$\Delta E=7.2\times 10^{-5}$~cm$^{-1}$. 
Analogously, the levels $\ket{ \nicefrac{3}{2}, \nicefrac{3}{2}, \nicefrac{3}{2},-1}$ and 
$\ket{ \nicefrac{3}{2}, \nicefrac{1}{2}, \nicefrac{3}{2},1}$ undergo two avoided intersections. 
The positions of 
these avoided intersections is shifted as $\beta$ is varied,  and their 
 widths  increase as $\beta$ increases and approaches $90\degree$.  
In the strong electric field regime, the energies of the states having the same field-free  $M_J$ run
parallel as $\Estatabs$ is augmented. 
For perpendicular fields, the Hamiltonian is invariant under  
the reflection in the plane perpendicular 
to the magnetic field. 
When the electric field is strong enough to overcome the $\Lambda$-doublet splitting,  then,  the 
levels having the same field-free $|M_J|$ form   
pairs of quasidegenerate  states.  For OH, the energy gap within these doublets decreases as 
$\Estatabs$ is increased. Whereas for LiO, again due to the small $\Lambda$-doubling, the two states
are quasidegenerate even for weak dc fields.

In the parallel field configuration, 
the states with field-free  odd and even parity are oriented and antioriented, respectively,
see ~\autoref{fig:fig_OH_B_fixed_cos_versus_Es}~(a) and 
\autoref{fig:fig_LiO_B_fixed_cos_versus_Es}~(a), but the additional magnetic field does not modify
significantly their absolute orientation $|\expected{\cos\theta_{\textup{s}}}|$. 
For LiO, $|\expected{\cos\theta_{\textup{s}}}|$ is larger for the state having a negative field-free
$M_J$.  
If the two fields are tilted, the presence of new avoided crossings modifies the directional
properties of the states, and $\expected{\cos\theta_{\textup{s}}}$
changes abruptly in the proximity of these irregular regions. 
Thus, for small variations of the electric field strength, 
a molecule prepared in the anti-oriented state would flip its dipole moment and turn into oriented. 
Such an electric field steered dipole switcher could have applications in tailoring the interactions between polar molecules.
By ramping the electric field strength through these crossings at different speeds,
the Landau-Zener dynamics could be studied 
and the Landau-Zener tunneling probability measured. 
For a strong enough electric field with $\beta>0\degree$, the pairs of states  
$\ket{ \nicefrac{3}{2}, M_J, \nicefrac{3}{2},\pm 1}$ have very close orientation. 
All the $J=3/2$ levels of LiO are orientated parallel to the field  in the strong dc-field regime, and
their $\expected{\cos\theta_{\textup{s}}}$ increases as $\Estatabs$ is increased.


\begin{figure}
\begin{center}
\includegraphics[width=\columnwidth]{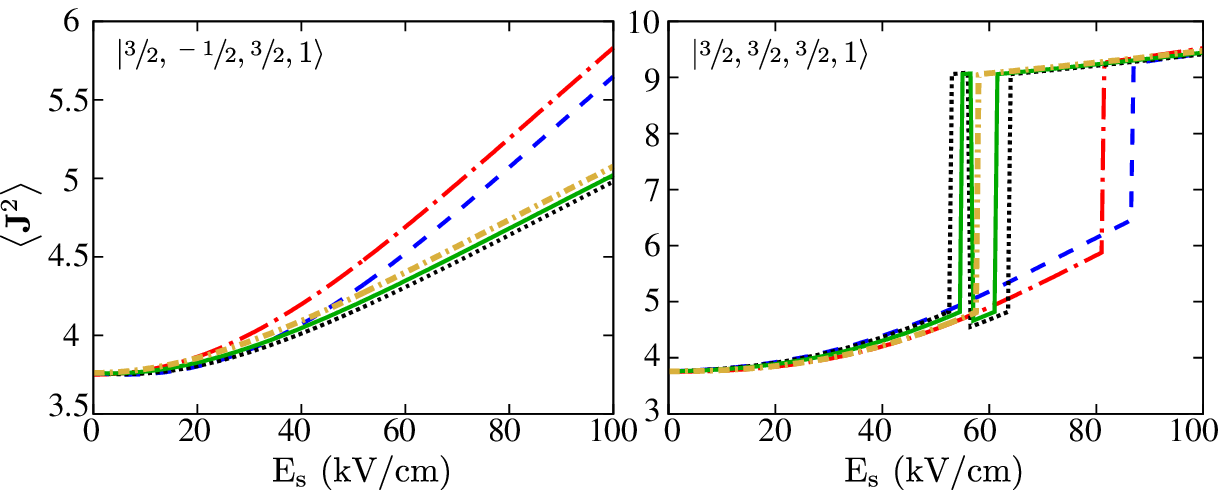}%
\caption{For LiO, we show the
expectation value $\expected{{\bf{J}}^2}$
of the states
$\ket{ \nicefrac{3}{2}, \nicefrac{-1}{2}, \nicefrac{3}{2},1}$ 
and 
$\ket{ \nicefrac{3}{2}, \nicefrac{3}{2}, \nicefrac{3}{2},1}$
 versus the electric field strength for 
$\Magabs= 0$~T (dot-long-dashed),
and 
$\Magabs= 1$~T 
with $\beta=0\degree$ (dashed),
$\beta=30\degree$  (dotted),
$\beta=60\degree$  (solid),
 and $\beta=90\degree$ (dot-short-dashed).}%
\label{fig:fig_LiO_B_fixed_j2_versus_Es}
\end{center}
\end{figure}

To illustrate the impact of an additional magnetic field on the  
hybridization of the angular motion of LiO, we have plotted 
the expectation value $\expected{{\bf{J}}^2}$ of 
$\ket{ \nicefrac{3}{2}, \nicefrac{-1}{2}, \nicefrac{3}{2},1}$ 
and 
$\ket{ \nicefrac{3}{2}, \nicefrac{3}{2}, \nicefrac{3}{2},1}$
as a function  of $\Estatabs$
in~\autoref{fig:fig_LiO_B_fixed_j2_versus_Es}~(a) and
\autoref{fig:fig_LiO_B_fixed_j2_versus_Es}~(b), respectively. 
There we provide a comparison between the results of the field configurations
$\Magabs= 1$~T with $\beta=0\degree$, $30\degree$ $60\degree$ and
$90\degree$ with those obtained with only an electric field.
As indicated above, the level $\ket{ \nicefrac{3}{2}, \nicefrac{-1}{2}, \nicefrac{3}{2},1}$ suffer real crossings with neighbouring levels 
for $\beta=0\degree$; and due to the level reordering 
it suffers an avoided crossings for $\beta>0\degree$ with a state in the same $J=3/2$ manifold.
The effect of this avoided crossing on $\expected{{\bf{J}}^2}$ is very small, 
because both levels have the same field-free value of $\expected{{\bf{J}}^2}$. 
Thus, independently of $\beta$, 
its $\expected{{\bf{J}}^2}$ shows a smooth and  increasing behaviour as $\Estatabs$ is increased. 
Furthermore, independently of $\Estatabs$, 
$\expected{{\bf{J}}^2}$ achieves the largest value in the absence of the magnetic field.
This is explained by the increase of 
the energy separation of this state from the neighbouring $J=5/2$ levels for $\beta>0\degree$,
which results in a reduction of the hybridization of the angular motion, and, therefore, of
$\expected{{\bf{J}}^2}$. 
A different behaviour is observed for the state $\ket{ \nicefrac{3}{2}, \nicefrac{3}{2}, \nicefrac{3}{2},1}$. 
For all these  field configurations,  
$\ket{ \nicefrac{3}{2}, \nicefrac{3}{2}, \nicefrac{3}{2},1}$  has the larger energy  within the $J=3/2$ manifold of $^2\Pi_{3/2}$ and 
shows a low-field seeking character when the dc field is turned on.
As a consequence, it suffers several avoided crossings with high-field seekers from the $J=5/2$ manifold of $^2\Pi_{3/2}$,
which are reflected as abrupt changes of $\expected{{\bf{J}}^2}$ when $\Estatabs$
is varied. Note that for $\beta=30\degree$ and $60\degree$, this state undergoes two consecutive avoided crossings.
A third kind of behaviour is encountered:
For levels such as the rotational ground state $\ket{ \nicefrac{3}{2}, \nicefrac{-3}{2}, \nicefrac{3}{2},-1}$,
the additional magnetic field provokes only minor variations on  $\expected{{\bf{J}}^2}$.  

The behavior for a weak  electric field illustrates nicely an interesting feature of open shell states
compared to $\Sigma$ electronic states: 
In the presence of only a magnetic field,
the ground state 
$\ket{ \nicefrac{3}{2}, \nicefrac{3}{2}, \nicefrac{3}{2},-1}$ 
is separated from 
$\ket{ \nicefrac{3}{2}, \nicefrac{3}{2}, \nicefrac{3}{2},1}$ 
by the $\Lambda$-doubling splitting.
By switching on an electric field, these two
almost degenerate states mix immediately, and the oriented superposition
becomes the ground state while the anti-oriented one 
increases in energy with $\Estatabs$ (see \autoref{fig:fig_OH_B_fixed_E_versus_Es} and
\autoref{fig:fig_LiO_B_fixed_E_versus_Es}). The two states of this doublet are indeed strongly
oriented and antioriented, respectively, before they start to interact with other states. If one
could prepare the ground state in such a field configuration, this amounts to a two-level molecule
with an oriented and an antioriented component.  
If the dc field is turned on slowly, the molecule adiabatically follows the oriented ground state.
Whereas, population could be transferred to the antioriented level if the switching on is done 
fast enough.
In a molecular ensemble, where dipole-dipole interaction has to be taken into account, this allows one
to access different phases of the many-body (spin-) system by tuning the field parameters or the speed
of the field ramps. The implementation of spin Hamiltonians with a high degree of control over the system parameters
may be achieved this way. Also, field parameters could be used to shape the intermolecular interactions
themselves~\cite{PhysRevLett.98.060404,PhysRevA.83.051402,1367-2630-14-4-043018}. 

\subsubsection{Constant electric field strength and increasing magnetic field strength}

\begin{figure}
\begin{center}
\includegraphics[width=\columnwidth]{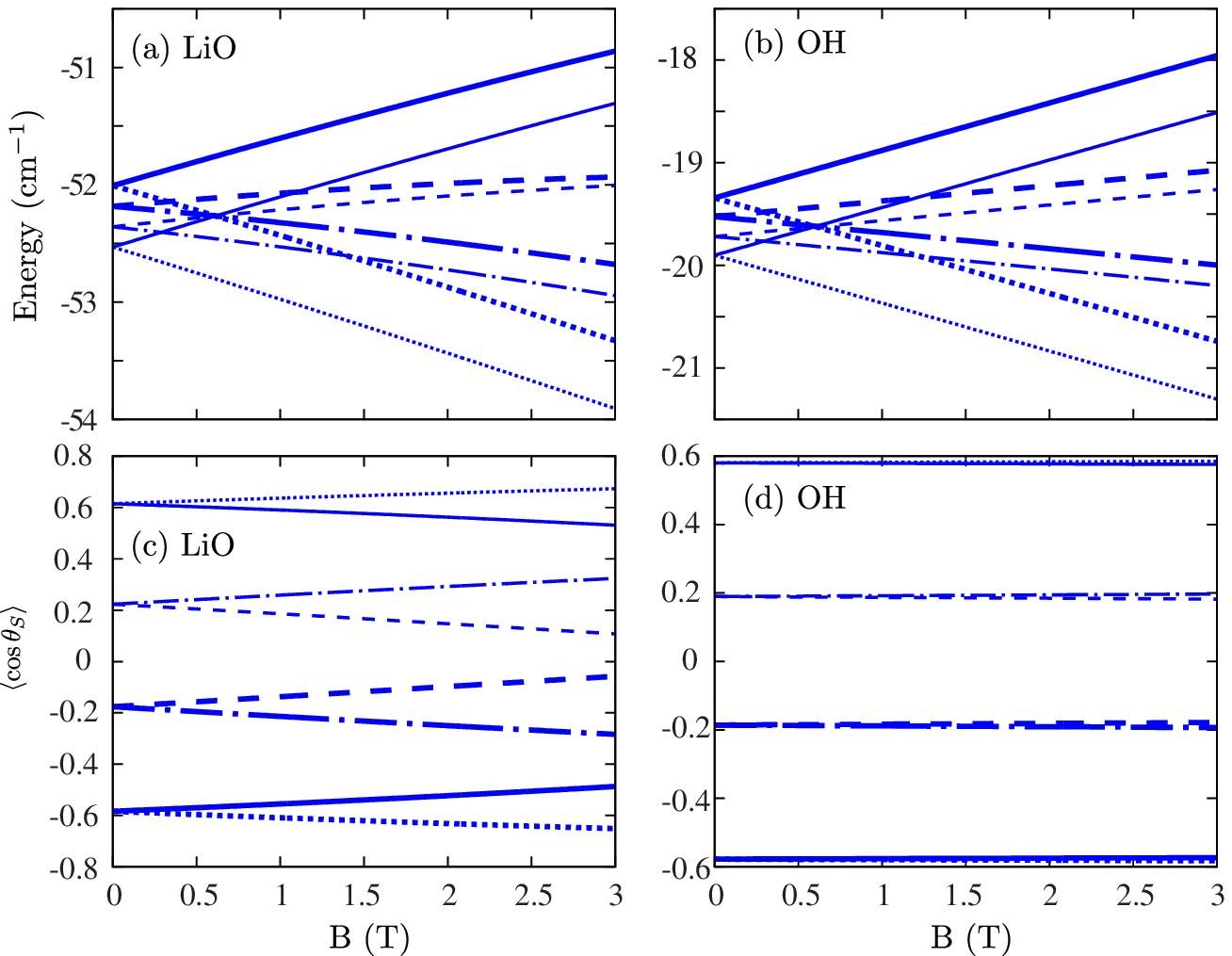}
\caption{For the $J=3/2$ levels of the $^2\Pi_{3/2}$ state
of LiO and OH,  
energy and expectation value $\expected{\cos\theta_{\textup{s}}}$ 
as a function of $\Magabs$ for the 
field configuration 
$\Estatabs= 4$~kV/cm and 
$\Estatabs= 17$~kV/cm, respectively
and $\beta=0\degree$.
The labeling of the states is the same as in \autoref{fig:fig_magnetic_field_olny}. 
}
\label{fig:impact_B_fixed_Es_LiO_OH_0}
\end{center}
\end{figure}

\begin{figure}
\begin{center}
\includegraphics[width=\columnwidth]{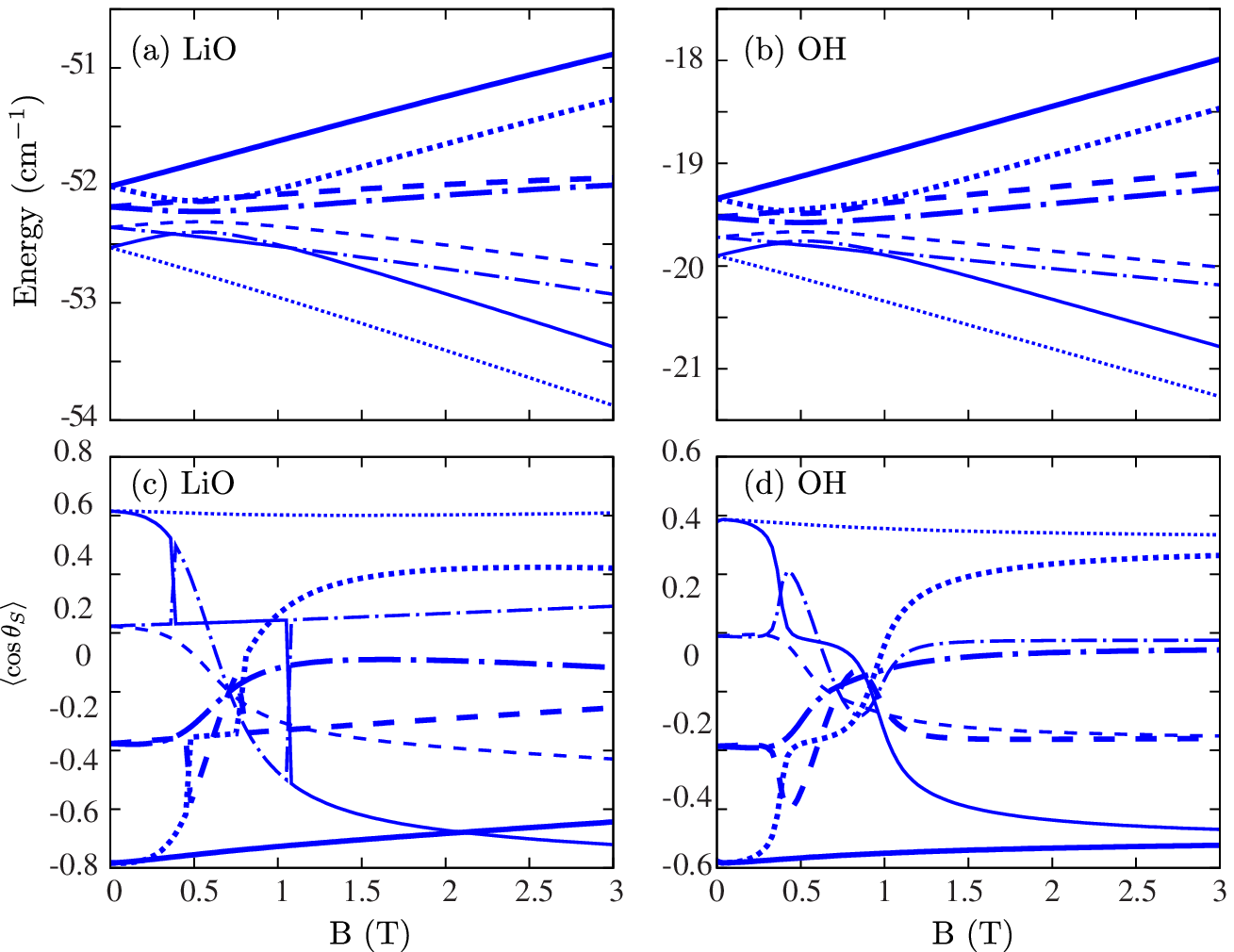}%
\caption{Same as \autoref{fig:impact_B_fixed_Es_LiO_OH_0}
but for $\beta=30\degree$}
\label{fig:impact_B_fixed_Es_LiO_OH_30}
\end{center}
\end{figure}

Here we analyze the impact of increasing the magnetic field strength, when the system is also exposed to an
electric field, which is turned on first.
For the $J=3/2$ levels of the LiO and OH $^2\Pi_{3/2}$ electronic states, 
\autoref{fig:impact_B_fixed_Es_LiO_OH_0}
and
\autoref{fig:impact_B_fixed_Es_LiO_OH_30} display the 
energies and orientation cosines as $\Magabs$ is increased 
for $\beta=0\degree$ and $\beta=30\degree$, respectively. 
For LiO and OH, the dc field strengths are fixed to $\Estatabs= 4$~kV/cm and
$\Estatabs= 17$~kV/cm, respectively, so that at 
$\Magabs= 1$~T and $\beta=0\degree$
the Zeeman and Stark interactions are of the same order of magnitude.

For parallel fields,  the field-free states with $s=1$ show a high-field seeking behaviour as
$\Magabs$ is increased, whereas those with $s=-1$ are low-field seekers.
Since $M_J$ is still a good quantum number, we encounter 
several exact crossings among these levels, whose positions are different for LiO and OH.
The  directional properties of OH are not affected by the additional magnetic field. 
In contrast, 
$\expected{\cos\theta_{\textup{s}}}$ increases (decreases) for the levels of LiO
with field-free even (odd) parity as  $\Magabs$ is enhanced.

When the fields are tilted, all the real intersections become avoided, and 
the complexity of the field-dressed spectrum increases significantly.  
The widths of these avoided crossings increases as $\beta$ is increased toward $90\degree$.
Compared to LiO, the avoided crossings in OH are broader. For instance, 
the OH levels 
$\ket{ \nicefrac{3}{2}, \nicefrac{+3}{2}, \nicefrac{3}{2},-1}$  and
$\ket{ \nicefrac{3}{2}, \nicefrac{-1}{2}, \nicefrac{3}{2},-1}$ suffer an avoided crossing 
at $\Magabs= 0.3778$~T with $\Delta E=1.386\times 10^{-2}$~cm$^{-1}$,
whereas for LiO it occurs at  $\Magabs= 0.3750$~T with $\Delta
E=9.4\times 10^{-5}$~cm$^{-1}$. 
The states $\ket{ \nicefrac{3}{2}, \nicefrac{-3}{2}, \nicefrac{3}{2},-1}$  and
$\ket{ \nicefrac{3}{2}, \nicefrac{3}{2}, \nicefrac{3}{2},1}$  do not suffer any avoided crossing, and their
orientation smoothly varies as  $\Magabs$ is increased. Compared to the parallel field
configuration, their absolute orientation $|\expected{\cos\theta_{\textup{s}}}|$
is reduced.
For the remainder of the states,  $\expected{\cos\theta_{\textup{s}}}$ changes abruptly 
in the proximity of the avoided crossing. Once the magnetic field is strong enough
their $\expected{\cos\theta_{\textup{s}}}$ also shows a smooth behaviour as a function of  $\Magabs$.

\subsubsection{Influence of the inclination of the fields}

\begin{figure}
\begin{center}
\includegraphics[width=\columnwidth]{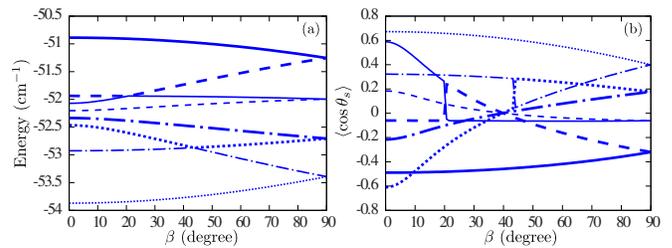}%
\caption{For the $J=3/2$ levels of the $^2\Pi_{3/2}$ state of LiO, (a) the energy
and (b) the expectation value $\expected{\cos\theta_{\textup{s}}}$ are plotted as a function
of the inclination angle $\beta$. 
The field configuration is $\Magabs= 2$~T  and 
$\Estatabs= 10$~kV/cm.
The states are labeled as in~\autoref{fig:fig_magnetic_field_olny}.}
\label{fig:impact_beta}
\end{center}
\end{figure}

In this section, we investigate  the impact of the inclination angle on the 
rotational dynamics. For the lowest lying rotational manifold of LiO, 
\ie, the $J=3/2$ levels of $^2\Pi_{3/2}$,  we
present in \autoref{fig:impact_beta}~(a) and 
\autoref{fig:impact_beta}~(b) the energy and 
orientation of the dipole moment with respect to the electric field,
$\expected{\cos\theta_{\textup{s}}}$, respectively, as a function of $\beta$ 
with $\Estatabs= 10$~kV/cm and $\Magabs= 2$~T.
In this regime of field strengths, 
only states within a rotational manifold suffer avoided crossings with each
other. The admixture of states having $J=5/2$ is very small.
For $\beta=0\degree$, the degeneracy of the states is lifted, with $M_J$ being still a good quantum number.
As $\beta$ is varied the symmetries of
the system are reduced,  and for orthogonal fields, pairs of quasidegenerate states are formed,
which is a sign for the emergence of a new symmetry.

The states  
$\ket{ \nicefrac{3}{2}, \nicefrac{-3}{2}, \nicefrac{3}{2}, -1}$,
$\ket{ \nicefrac{3}{2},  \nicefrac{3}{2},\nicefrac{3}{2},  1}$,
$\ket{ \nicefrac{3}{2}, \nicefrac{-1}{2},\nicefrac{3}{2},  1}$, and
$\ket{ \nicefrac{3}{2},  \nicefrac{1}{2},\nicefrac{3}{2}, -1}$
do not suffer any avoided crossings as $\beta$ is varied, and their orientation changes smoothly. 
The state $\ket{ \nicefrac{3}{2}, \nicefrac{-3}{2},  \nicefrac{3}{2},  -1}$ presents the larger orientation, and 
$\expected{\cos\theta_{\textup{s}}}$ is reduced from $0.67$ for $\beta=0\degree$ 
till $0.40$ for $\beta=90\degree$. 

The pair $\ket{  \nicefrac{3}{2},  \nicefrac{-3}{2}, \nicefrac{3}{2},   1}$ and 
$\ket{   \nicefrac{3}{2}, \nicefrac{-1}{2},  \nicefrac{3}{2},  -1}$ suffer an avoided
crossing, and around this irregular region their orientation changes abruptly, analogously for  the 
$\ket{  \nicefrac{3}{2},  \nicefrac{3}{2},  \nicefrac{3}{2},  -1}$ and $\ket{ \nicefrac{3}{2}, \nicefrac{1}{2}, \nicefrac{3}{2},   1}$ levels. 
After these avoided crossings,
their directional properties evolve smoothly as $\beta$ is enhanced towards $90\degree$.
For $\beta=39.5\degree$, the states 
$\ket{  \nicefrac{3}{2}, \nicefrac{-3}{2}, \nicefrac{3}{2},  1}$, 
$\ket{  \nicefrac{3}{2}, \nicefrac{-1}{2}, \nicefrac{3}{2},  1}$, 
$\ket{  \nicefrac{3}{2},  \nicefrac{1}{2}, \nicefrac{3}{2},  1}$, and
$\ket{  \nicefrac{3}{2},  \nicefrac{1}{2}, \nicefrac{3}{2},  -1}$ 
do not show any orientation.

\subsection{A toy molecular system}
In this section, we explore the possibility of using external fields
to couple rotational levels from the two 
fine structure components of a $^2\Pi$ electronic state. 
Our  aim  is to investigate field-induced couplings between spin degrees
of freedom and the molecular rotation. 
We are interested in predicting general properties and  principal effects  with
the focus on their understanding, thus we do not address a specific
molecule but use a toy system. 
This molecular model has been constructed using the  OH
molecule as a prototype: its rotational and spin-orbit constants have been reduced by a factor of $10$, whereas
the spin-rotation and $\Lambda$-doubling parameters by a factor of $100$. We have used this
ratio based on the simple pure precession hypothesis to estimate the $\Lambda$-doublet parameter 
$p=4AB/\Delta E$, where $\Delta E$ is the energy separation between the
$^2\Pi$ electronic ground state and the $^2\Sigma$ lowest excited electronic state. The dipole moment has been fixed to 
$\mu=2$~D. Note that both, increasing the dipole moment or decreasing the rotational or spin-orbit constant, 
lead to the appearance of crossings at lower fields. 

For this system, the field-free spectrum is similar to the OH energy structure presented in~\autoref{fig:fig_1},
but the energy scales have been reduced so that the coupling between different degrees of freedom can be achieved at 
lower field strengths. 
In the $^2\Pi_{3/2}$ fine structure component, the  $J=3/2$ and $5/2$ rotational bands are 
separated by
$8.4$~cm$^{-1}$, and they have $\Lambda$-doublet splittings of $5.6\times 10^{-4}$~cm$^{-1}$  and
$2\times 10^{-3}$~cm$^{-1}$, respectively. 
The ground state of  $^2\Pi_{1/2}$ lies  $12.6$~cm$^{-1}$ and $4.2$~cm$^{-1}$  above
the $J=3/2$ and $J=5/2$ levels of $^2\Pi_{3/2}$, respectively, and 
has a $\Lambda$-doublet splitting of $1.6\times 10^{-3}$~cm$^{-1}$. 

\begin{figure}[t]
\includegraphics[width=\columnwidth]{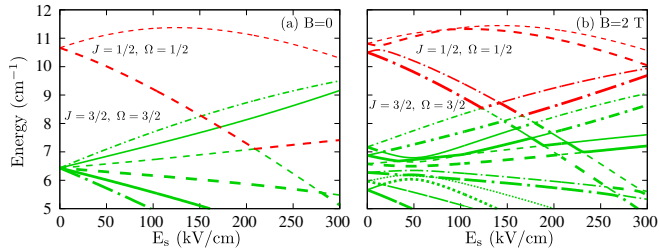}
\caption{For the toy system, we show the energy of the states with 
  $J=5/2$ (green) from $^2\Pi_{3/2}$ and
  $J=1/2$ (red) 
from $^2\Pi_{1/2}$ as a 
function of the dc field strength $\Estatabs$ for
the field configuration 
(a) $\Magabs= 0$~T  and  $\beta=0\degree$
and 
(b) $\Magabs= 2$~T  and  $\beta=30\degree$.
The labels of states are:
$M_J=1/2$ and $s=\pm1$ thick and thin dashed, 
$M_J=-1/2$ and $s=\pm1$ thick and thin dot-dashed,
$M_J=3/2$ and $s=\pm1$ thick and thin solid, 
$M_J=-3/2$ and $s=\pm1$ thick and thin dotted, 
$M_J=5/2$ and $s=\pm1$ thick and thin dot-short-dashed,
and
$M_J=-5/2$ and $s=\pm1$ thick and thin long-dashed.
}%
\label{fig:toy_model_impact_static}
\end{figure}

\autoref{fig:toy_model_impact_static}~(a) displays the energies for the levels with
$J=3/2$ and $J=5/2$ of $^2\Pi_{3/2}$ and $J=1/2$ of $^2\Pi_{1/2}$ as a function of the 
electric strength. When the dc field is turned on, the levels in a $J$-manifold are
driven apart in pairs having the same $|M_J|$ and depending on the field-free parity they
have a high- or low-field seeking character.
The $\ket{ \nicefrac{1}{2}, \nicefrac{\pm1}{2},  \nicefrac{1}{2},   1}$ levels are high-field seekers, then, when $\Estatabs$
is increased they will encounter the low-field seeking states from the $J=5/2$ manifold of $^2\Pi_{3/2}$.
Indeed, 
 $\ket{  \nicefrac{1}{2}, \nicefrac{\pm1}{2},  \nicefrac{1}{2},  1}$  undergoes exact crossings with $\ket{  \nicefrac{5}{2}, \nicefrac{\pm5}{2},  \nicefrac{3}{2},-1}$
and  $\ket{  \nicefrac{5}{2}, \nicefrac{\pm3}{2},  \nicefrac{3}{2}, -1}$  
for $\Estatabs= 151.698$~kV/cm and $\Estatabs= 171.718$~kV/cm, respectively.
In addition, it suffers a very narrow avoided crossing with 
$\ket{   \nicefrac{5}{2}, \nicefrac{\pm1}{2}, \nicefrac{3}{2}, -1}$  
for $\Estatabs\approx 209.22$~kV/cm 
of  width $\Delta E=2.1\times  10^{-5}$~cm$^{-1}$, 
 and through it these levels interchange their intrinsic
character. 
At these irregular regions, the dc field induces a strong mixing and interaction between the rotational and spin degrees of freedom. 
Thus, we encounter interactions between states in different rotational manifolds of given 
fine structure components, but also among levels in $^2\Pi_{3/2}$ and $^2\Pi_{1/2}$. 

The results for this molecule exposed to an additional
 magnetic field of $2$~T forming an angle $\beta=30\degree$ with the dc field 
 are presented in \autoref{fig:toy_model_impact_static}~(b).
Now, $M_J$ is not a good 
quantum number, the $|M_J|$-degeneracy is lifted and 
the real intersections among levels appearing in  an electric field become avoided. Thus,
the first avoided crossing between $^2\Pi_{3/2}$ and $^2\Pi_{1/2}$ states takes place at a lower
electric field strength 
$\Estatabs$. 
Indeed, the states
$\ket{  \nicefrac{1}{2}, \nicefrac{-1}{2},  \nicefrac{1}{2},  1}$ 
and
$\ket{  \nicefrac{5}{2}, \nicefrac{5}{2},  \nicefrac{3}{2},-1}$
suffer an avoided intersection for $\Estatabs\approx 122.898$~kV/cm, and energy 
width  $2.96 \times 10^{-4}$~cm$^{-1}$.
Thus the  effects due to the mixing of different fine structure components come
into reach under  experimentally realistic conditions. 
By further increasing $\Estatabs$, we encounter a 
cascade of avoided  crossings within the $J=5/2$ rotational manifold of $^2\Pi_{3/2}$.
 We have identified the avoided crossing equivalent to the one at
  $ 209.22$~kV/cm in the absence of the magnetic field. It  now takes place between the states
  $\ket{\nicefrac{1}{2},\nicefrac{-1}{2},\nicefrac{1}{2},1}$ and
  $\ket{\nicefrac{5}{2},\nicefrac{1}{2},\nicefrac{3}{2},1}$ for
  $\Estatabs= 204.474$~kV/cm and has a larger energy gap 
$\Delta E=1.4\times 10^{-4}$~cm$^{-1}$.

\section{Conclusions}
\label{sec:conclusions}
We have investigated the impact of combined
electric and magnetic fields on the rotational spectrum of 
open shell diatomic molecules in a $^2\Pi$ electronic state.
This study has been performed within the rigid rotor approximation, including the fine-structure 
interactions and the $\Lambda$-doubling effects. 
For several field configurations, the richness and variety of the  field-dressed
rotational dynamics has been illustrated by analyzing the
energies, the directional properties, and the hybridization of the angular motion.

Considering several field regimes, we have explored the role of the different interactions, and the 
possibility of inducing coupling between the different degrees of freedom by using the external fields.
Due to the large rotational constant of the OH radical, we have shown that 
for the $J=3/2$ states of $^2\Pi_{3/2}$, the field-dressed dynamics takes place within this manifold
even for strong external fields, and the contribution of levels with higher $J$ is negligible.
In contrast, we have proven that strong electric or magnetic fields induce coupling  between states of different rotational 
manifolds of $^2\Pi_{3/2}$ in the LiO spectrum. 
Finally, we have considered a realistic model system and demonstrated the feasibility of inducing coupling between rotational levels
of the $^2\Pi_{1/2}$ and $^2\Pi_{3/2}$ electronic states by means of external fields. 

In all these regimes, the complexity of the field-dressed spectrum is characterized by the amount of
avoided crossings that the states suffer in the  field-dressed dynamics. 
Around them, the fields become control knobs that could be exploited to tailor the interactions between polar molecules 
and influence their rotational dynamics.
Indeed, by  tuning the field parameters one can find configurations where 
either  the dipole moment  and orientation are flipped or the electronic spin interchanged. 
Then, Landau-Zener tunnelling between these states could be observed if the corresponding field 
strength is changed fast enough around the crossing region, and Rabi oscillations could be induced.
Here, we have shown that these phenomena might appear at field strengths which are well within experimental reach.

\begin{acknowledgments}

Financial support by the Spanish project FIS2011-24540 (MICINN), the
Grants P11-FQM-7276 and FQM-4643 (Junta de Andaluc\'{\i}a), and the Andalusian research group FQM-207 is
gratefully appreciated. J.J.O. acknowledges the support of ME under the program FPU.

\end{acknowledgments}


\end{document}